# Waves in a short cable at low frequencies, or just hand-waving?

What does physics say?

(Invited paper)

L.B. Kish

Department of Electrical Engineering
Texas A&M University
College Station, TX 77843-3128, USA

S.P. Chen

Department of Electrical Engineering
Texas A&M University
College Station, TX 77843-3128, USA

C.G. Granqvist

Department of Engineering Sciences
The Ångström Laboratory, Uppsala University
P.O. Box 534, SE-75121 Uppsala, Sweden

J.M. Smulko

Department of Metrology and Optoelectronics
Faculty of Electronics, Telecommunications and Informatics
Gdansk University of Technology
Narutowicza 11/12, 80-233 Gdansk, Poland

*Abstract*—We address the question of low-frequency signals in a short cable, which are often considered as waves in engineering calculations. Such an assumption violates several laws of physics, but exact calculations can be carried out via linear network theory.

I. INTRODUCTION

Recently, a, seemingly effective, new attack [1] against the Kirchhoff-law–Johnson-noise (KLJN) secure key exchange scheme [2,3] was published. Even though the KLJN scheme has proven unconditional security [3], the claimed strong information leak, if valid, would have severely limited its practical applications. Soon it was shown that the attack [1] had both theoretical [4] and experimental [5] flaws and that it is no more efficient than the old wire resistance based attacks [6]. Moreover, even this negligible information leak can be nullified by a recent defense protocol [6].

While the new attack [1] became obsolete, an interesting question, which is unrelated to security, remained. The authors behind the new attack [1] used a wave picture to deduce their results. However, the KLJN scheme operates in the quasi-static limit of electrodynamics, which means that the cable length is much shorter than the wavelength. In a physical description of cables with finite length $D$, the frequency-space of wave solutions is quantized to discrete values so that integer multiples of the half-wavelength fit in the cable. Thus the longest allowed wavelength $\lambda_{max}$ of the wave, and the lowest allowed frequency $f_{min}$ at phase velocity $v$, can be written as

$$\lambda_{max} = 2D \quad , \quad f_{min} = \frac{v}{2D} \; , \tag{1}$$

respectively. Frequencies below $f_{min}$, down to zero frequency, constitute a forbidden band of wave states. The KLJN condition for the signal (noise) bandwidth $B$ is

$$B << \frac{v}{2D} \; , \tag{2}$$

which excludes the possibility of wave solutions in the band. This is how physicists treat this problem.

However, we have learned from electrical- and microwave engineers that, during everyday practice, such low-frequency signals in small systems are usually treated as waves and that those wave-based mathematical solutions are taken to be exact.

A recent paper of ours [7] clarified the situation. Here we summarize those arguments and also show how an engineering-mathematical calculation can be exact in its final result but, at the same time, unphysical in its internal steps. The final conclusion is that the assumption of waves in the low-frequency (quasi-static) regime violates several laws of physics, yet it is fine to use waves in engineering calculations provided the waves in these internal steps are not considered physically existent. The observed propagation delays are not physical waves but *retarded potentials* without significant involvement of the conjugate physical quantities, which would play an identical role in a real electromagnetic wave.

## II. WAVE SOLUTIONS IN LINEAR RESPONSE THEORY

In linear time-invariant network analysis, the weighting function $h(t)$ is the response to a Dirac pulse at the input, as shown in Fig. 1.

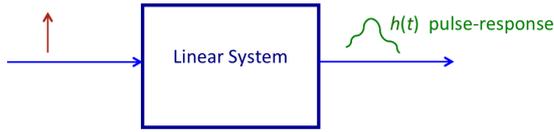

Fig. 1. The weighting function $h(t)$ is the response to a Dirac pulse at the input.

Any signal $x(t)$ with limited bandwidth at the input is then divided into abstract zero-width pulses, and the time response to these input pulses at the output are calculated and summed up by a convolution to get the output signal $Y(t)$; see Fig. 2.

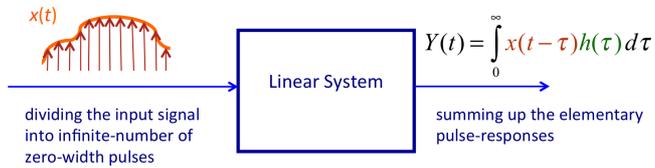

Fig. 2. The output signal $Y(t)$ is the convolution of elementary responses due to the abstract input pulses with zero width and infinite bandwidth.

The zero-width input pulses have infinite bandwidth, even though the total input signal may be very slow. Thus, if the linear system is a short cable and the signal bandwidth satisfies Rel. 2, the zero-width input pulses still satisfy Eq. (1) (excepting an infinitesimally small fraction of their Fourier spectrum). These pulses will hence propagate through, and reflect in, the cable as waves. The propagation can be described by waves and at the end, during the output summation, all wave components disappear as a consequence of their destructive interference so that we regain the slow, non-wave output signal $Y(t)$.

It is obvious that these abstract pulses and their propagation are non-physical because, for example, we cannot detect X-rays or impacts of the uncertainty principle due to the high bandwidth and short time scales involved. In engineering and physics, many similar mathematical contrivances are used to calculate results in an easy way. Clearly, mathematics is a much broader field than physics, and physics will ultimately make use of those mathematical solutions that are physical. It is important not to assign physical characteristics to elements that are only non-physical mathematical constructs.

## III. PROPAGATION OF RETARDED NON-WAVE POTENTIALS

### A. What is a physical wave, and what makes it different from a propagating slow, non-wave signal?

(*i*) A physical wave is a dynamical oscillation that takes place via the propagation in space without energy decay in a loss-free medium and without the need of an external generator to keep it going.

(*ii*) It is also essential that there are two oscillating, dual energy forms (with conjugate physical quantities), and the total energy is the sum of these two energies, which are equally distributed. These energies are manifest via a dynamical transfer into the dual form, and back, during propagation: they "induce" and "regenerate" each other during the propagation without loss (in loss-free media).

*Example 1*: In the case of electromagnetic waves, the two energy forms are electric and magnetic, and the way of regenerating each other is via induction and displacement currents.

*Example 2*: In the case of elastic waves, the two energy forms are potential energy due to deformations and kinetic energy due to motion, and the way of regenerating each other is via Newton's and Hook's laws.

It is important that, in a loss-free medium, this energy transfer must have 100% efficiency since otherwise the wave would decay during propagation (without external generator drive) and then it is not an electromagnetic wave but a non-wave-type retarded potential, such as the near-field around an antenna or the magnetic and electrical field around a wire.

Figures 3 and 4 show two situations where travelling structures (or their fields) look like waves for the superficial observer who does not execute a deeper investigation toward the nature of these propagating oscillations.

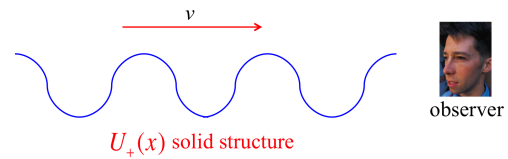

Fig. 3. Non-wave-type retarded oscillation: a wave-shaped solid structure (such as wire) is traveling with velocity $v$ in positive direction along the $x$ axis. The observer sees it as a propagating wave. While the oscillation satisfies d'Alembert's equation for waves, this is not a wave. The dynamical transfer between the two dual energy forms is missing.

Surprisingly, both non-wave oscillations in Figs. 3 and 4 satisfy d'Alembert's equation [1] typically used for waves, *i.e.*,

$$U(t,x) = U_+\left(t - \frac{x}{v}\right), \quad (3)$$

thus indicating that d'Alembert's equation is not necessarily for waves but for propagating oscillations in general.

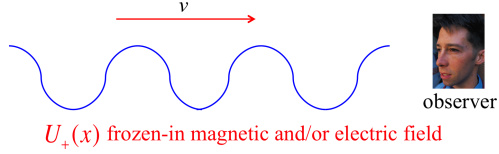

Fig. 4. Non-wave-type retarded oscillation: a wave-shaped frozen-in magnetic field or electrical field structure (such as in a magnetic tape or electret tape) is traveling with velocity $v$ in positive direction along the $x$ axis. The observer sees it as a propagating wave. While the oscillation satisfies d'Alembert's equation for waves, this is not a wave. The dynamical transfer between the two dual energy forms is missing.

### B. Electric and magnetic energy balance in a driven short cable at low frequencies

Consider now a short cable satisfying Rel. 2, which is driven by an ac voltage generator $U(t)$ at one end and, at the other end, is terminated with the wave resistance $R_w$ given as

$$R_w = \sqrt{\frac{L_u}{C_u}}, \qquad (4)$$

where $L_u$ and $C_u$ are the unit length (1 meter) cable inductance and cable capacitance, respectively. Then the cable has equal amounts of energy in the electric and magnetic forms, as discussed below, and both of these energies are oscillating. However, these energies are not "bouncing" between the dual electric and magnetic forms but oscillate separately between these forms and the generator. This fact is obvious from the considerations that follow next.

Quasi-static conditions (Rel. 2) imply that the time-dependent voltage and current are spatially homogeneous along the wire, *i.e.*,

$$U(x,t) \simeq U(t), \qquad (5)$$

$$I(x,t) \simeq I(t). \qquad (6)$$

Thus the electrical and magnetic energies, denoted $E_e(t)$ and $E_m(t)$, respectively, in the cable can be written

$$I(x,t) \simeq I(t), \qquad (7)$$

$$E_m(t) = \frac{1}{2} L_c I^2(t) = \frac{1}{2} L_c \frac{U^2(t)}{R_w^2}, \qquad (8)$$

where $C_c = DC_u$ and $L_c = DL_u$ are the capacitance and inductance of the whole cable, respectively, and the last term of Eq. (8) originates from the well-known fact that a cable of arbitrary length, closed with the wave impedance (resistance) $R_w$, has an input impedance of exactly $R_w$. Thus

$$\frac{E_e(t)}{E_m(t)} = \frac{C_c}{L_c} R_w^2 = \frac{C_u}{L_u} R_w^2 = 1 \qquad (9)$$

at each moment. Obviously, no "bouncing" phenomenon takes place between the dual electric and magnetic forms of energy, because the energy is bouncing rather between the generator and the magnetic and electric fields.

Suppose now that $R_A$ and $R_B$, rather than $R_w$, terminate the cable ends. The only significant change that will occur in the quasi-static limit is that the ratio of electrical and magnetic energies will differ from unity, which further goes against the wave hypothesis.

In conclusion, even the simple picture of wave modes given above proves that waves do not exist in a cable in the quasi-static limit.

## IV. SHORT CABLE DRIVEN BY LOW-FREQUENCY JOHNSON NOISE

Simple but fundamental thermodynamic considerations also prove that waves cannot exist in a cable in the quasi-static frequency range, as elaborated below.

### A. Proof that the total energy of a cable is less than the energy required for a single wave mode

As an example, we mathematically analyze how the no-wave situation manifests itself in a lossless cable, which is closed at both ends by resistors equal to the wave impedance $R_w$ at the temperature $T$. These conditions are not necessary but serve to simplify the calculations. Suppose that, in accordance with Rel. 2, the Johnson noise of the resistors has an upper cut-off frequency at $f_c$ so that

$$f_c \ll f_{min} = \frac{v}{2D} \qquad (10)$$

holds in order to satisfy Rel. 2. Thus partial homogeneity of current and voltage is valid for the cable, and it is straightforward to calculate the electric and magnetic energies of thermal origin, as shown next.

According to the Johnson–Nyquist formula, the thermal electrical energy in the cable capacitance is

$$E_{e,th} = \frac{1}{2} C_c \langle U^2(t) \rangle = \frac{1}{2} C_c \int_0^{f_c} \frac{4kTR_w/2}{1+f^2/f_{0C}^2} df \cong$$
$$\cong kTC_c R_w f_c = \frac{kT}{2} \frac{f_c}{f_{min}} \ll \frac{kT}{2} \qquad (11)$$

and the thermal magnetic energy in the cable inductance is

$$E_{m,th} = \frac{1}{2} L_c \langle I^2(t) \rangle = \frac{1}{2} L_c \int_0^{f_c} \frac{4kT/(2R_w)}{1+f^2/f_{0L}^2} df \cong$$
$$\cong kT \frac{L_c}{R_w} f_c = \frac{kT}{2} \frac{f_c}{f_{min}} \ll \frac{kT}{2} \qquad (12)$$

where the characteristic frequencies of the Lorentzian spectra in Eqs. (11) and (12) are defined as

$$f_{0C} = \frac{1}{2\pi C_c R_w /2} = \frac{1}{\pi D C_u}\sqrt{\frac{C_u}{L_u}} = \frac{1}{\pi D}\sqrt{\frac{1}{L_u C_u}} =$$
$$= \frac{1}{\pi}\frac{v_c}{D} = \frac{2}{\pi}f_{min} \qquad (13)$$

and

$$f_{0L} = \frac{2R_w}{2\pi L_c} = \frac{1}{\pi L_c}\sqrt{\frac{L_u}{C_u}} = \frac{1}{\pi D}\sqrt{\frac{1}{L_u C_u}} = \frac{1}{\pi}\frac{v_c}{D} = \frac{2}{\pi}f_{min} \ . \qquad (14)$$

Similar calculations can be carried out for the general case in which the cable ends are not terminated by $R_w$ but with different resistance values $R_A$ and $R_B$. Specifically, the parallel resultant resistance $R_A$ and $R_B$ (instead of $R_w/2$) enters in the left-hand side of Eqs. (11) and (13), and the serial (sum) loop resistance $R_A + R_B$ (instead of $2R_w$) enters in the left-hand side of Eqs. (12) and (14), while the final inequalities shown by Eqs. (11) and (12) remain.

If there is loss in the cable, and the cable is terminated by $R_A$ and $R_B$, the same situation holds provided $R_A$ and $R_B$ as well as the cable have the same (noise) temperature, because the system is still in thermal equilibrium. If the cable is cooler, then there is an energy flow out of the cable, which further strengthens the inequalities at the right-hand sides of Eqs. (11) and (12).

In conclusion, Eqs. (11) and (12) prove that for a short cable and within the frequency range of interest for the KLJN scheme, the sum of electrical and magnetic energies in all of the hypothetical "wave modes" is much less than the energy needed for a single wave mode in thermal equilibrium.

### B. Violation of the Energy Equipartition Theorem, the Principle of Detailed Balance and the Second Law of Thermodynamics

According to Boltzmann's Energy Equipartition Theorem [7] for thermal equilibrium at temperature $T$, each electromagnetic wave mode has a mean thermal energy equal to $kT$, where $k$ is Boltzmann's constant. Half of this mean energy is electrical and the other half is magnetic. For $N$ different wave modes in the system, both the electrical and magnetic fields carry a total energy equal to $NkT/2$.

It is easy to see that the assumption that, in a *hypothetical* wave system, these wave energies are less than the above given values violates not only the Energy Equipartition Theorem but also the Principle of Detailed Balance [6] and the Second Law of Thermodynamics: Coupling this hypothetical wave system to a regular one would hit the Detailed Energy Balance of equilibrium between wave modes in the hypothetical and regular systems because it would yield an energy flow toward the hypothetical one. This energy flow could then be utilized for perpetual motion machines of the second kind, *i.e.*, violate the Second Law of Thermodynamics.

### C. Violation of Planck's Law and experimental facts for blackbody radiation

Planck [7] deduced his law of thermal radiation, for simplicity, from the properties of a box with black walls, *i.e.*, internal walls with unity absorptivity and emissivity. In thermal equilibrium, thermal radiation within an infinitely large box with walls of arbitrary absorptivity, emissivity and color has a power spectral intensity for each polarization given by

$$I(f) = \frac{4\pi h f^3}{c^2}\frac{1}{e^{hf/kT}-1} \ , \qquad (15)$$

where $h$ is Planck's constant. The derivation of this formula is based on counting existing wave modes in the frequency range $f > f_{min}$, where the minimum frequency is obtained by the same frequency quantization as we claim exists in a cable. A certain misconception exists that, in closing the finite-size cable by the wave impedance (resistance) $R_w$ at its two ends, all of the lower frequencies, at $f < f_{min}$, will also be available for wave modes as a consequence of the unity absorptivity and emissivity of the impedance match. However, one must realize that a cable closed by the wave impedance (resistance) $R_w$ at its two ends is a one-dimensional realization of Planck's box with perfectly absorbing (black) walls. Allowing the $f < f_{min}$ frequency range for wave modes results in a non-quantized continuum distribution of wave modes, which yields an infinite number $N$ of wave modes in any finite frequency band. Such a situation in the finite-size cable results in an infinite amount of thermal energy $NkT$ in any finite frequency band within the classical-physical frequency range $f << kT/h$. This situation is not a problem in a cable or box with infinite size. However, in a finite-size box, infinite thermal energy in the sum of wave modes would naturally result in infinite intensity of blackbody radiation.

Furthermore, the situation is similar when the cable ends are terminated by $R_A$ and $R_B$ rather than by $R_w$. According to Planck's results, discussed above, the thermal radiation field in the closed box and cable does not depend on the absorptivity and emissivity of the walls—*i.e.*, on the cable termination resistances—since otherwise Planck's Law would be invalid and his formula violate the Second Law of Thermodynamics. Therefore the above argumentation regarding infinite energies and infinite thermal radiation holds for arbitrary termination and wall color.

If there is loss in the cable and the cable is terminated by $R_A$ and $R_B$, the same situation holds provided $R_A$, $R_B$ and the cable have the same (noise) temperature, because the system is still in thermal equilibrium.

In conclusion, assuming waves in the quasi-static limit (Rel. 2) violates Planck's Law and experimental facts about blackbody radiation.

## V. Conclusions

We discussed how non-physical elements used for calculation in engineering can lead to physical results, and why the non-physical elements should not be considered physical. We defined what physical waves are and showed that low-frequency signal propagation in a short cable does not entail waves but retarded potentials. Finally, we proved that the assumption of waves in the quasi-static limit violates several laws of physics, including (*i*) the Energy Equipartition Theorem, (*ii*) the Second Law of Thermodynamics (*i.e.*, allowing the construction of a perpetual motion machine of the second kind, (*iii*) the Principle of Detailed Balance, and (*iv*) Planck's formula at the low-frequency end (which would permit infinitely strong black body radiation there).